\documentclass[preprint,12pt]{elsarticle}
\makeatletter
\def\ps@pprintTitle{}
\makeatother

\usepackage{amssymb}
\usepackage{xcolor}
\usepackage{verbatim}
\usepackage{booktabs}
\usepackage{hyperref} 
\usepackage{multicol} 
\usepackage{multirow} 
\usepackage{subfig}
\usepackage{amsmath}
\usepackage{color, colortbl}
\usepackage{float}
	
\definecolor{Gray}{gray}{0.9}
\definecolor{Lavender}{RGB}{230, 230, 250}
\definecolor{Peach}{RGB}{255, 218, 185}
\definecolor{Mint}{RGB}{152, 255, 152}
\definecolor{Butter}{RGB}{255, 255, 153}

\begin{document}

\begin{frontmatter}



\title{Mitigating subjectivity and bias in AI development indices: A robust approach to redefining country rankings}

\author[inst1]{Betania Silva Carneiro Campello}

\affiliation[inst1]{organization={School of Applied Sciences},
            city={University of Campinas, Campinas},
            postcode={13083-970}, 
            state={São Paulo},
            country={Brazil}}

\author[inst2]{Guilherme Dean Pelegrina}

\author[inst2]{Renata Pelissari}

\affiliation[inst2]{organization={Mackenzie Presbyterian University},
            city={São Paulo},
            postcode={13083-970}, 
            state={São Paulo},
            country={Brazil}}

\author[inst3]{Ricardo Suyama}

\affiliation[inst3]{organization={Center for Engineering, Modeling and Applied Social},
            city={Federal University of ABC, Santo André},
            postcode={ 09280-560}, 
            state={São Paulo},
            country={Brazil}}

    \author[inst1]{Leonardo Tomazeli Duarte}

\fntext[myfootnote]{Work supported by São Paulo Research Foundation (FAPESP) under the grants \#2020/09838-0 (BI0S - Brazilian Institute of Data Science), \#2020/10572-5 and \#2023/04159-6. }

\begin{abstract}
 
Countries worldwide have been implementing different actions national strategies for Artificial Intelligence (AI) to shape policy priorities and guide their development concerning AI. Several AI indices have emerged to assess countries' progress in AI development, aiding decision-making on investments and policy choices. Typically, these indices combine multiple indicators using linear additive methods such as weighted sums, although they are limited in their ability to account for interactions among indicators. Another limitation concerns the use of deterministic weights, which can be perceived as subjective and vulnerable to debate and scrutiny, especially by nations that feel disadvantaged. Aiming at mitigating these problems, we conduct a methodological analysis to derive AI indices based on multiple criteria decision analysis. Initially, we assess correlations between different AI dimensions and employ the Choquet integral to model them. Thus, we apply the Stochastic Multicriteria Acceptability Analysis (SMAA) to conduct a sensitivity analysis using both weighted sum and Choquet integral in order to evaluate the stability of the indices with regard the weights. Finally, we introduce a novel ranking methodology based on SMAA, which considers several sets of weights to derive the ranking of countries. As a result, instead of using predefined weights, in the proposed approach, the ranking is achieved based on the probabilities of countries in occupying a specific position. In the computational analysis, we utilize the data employed in The Global AI Index proposed by Tortoise. Results reveal correlations in the data, and our approach effectively mitigates bias. In the sensitivity analysis, we scrutinize changes in the ranking resulting from weight adjustments. We demonstrate that our proposal rankings closely align with those derived from weight variations, proving to be more robust.

\end{abstract}




\end{frontmatter}


\section{Introduction}
\label{sec:int}

     Artificial Intelligence (AI) is a field focused on creating computer systems capable of executing tasks that usually demand human intelligence, including problem-solving, learning, and decision-making. AI has a high potential of transforming our society in profound ways, with implications for ethics, employment, privacy, and its impact extends across numerous domains, such as healthcare, finance, education, manufacturing, transportation, agriculture \cite{JAVAID202315, ALLGAIER2023102616, GOODELL2021100577}. Given this scenario, countries worldwide are increasingly recognizing the importance of developing national strategies and policies for its improvement \cite{WILSON2022101652, FATIMA2020178}.  
    
     These strategies focus on--to varying degrees depending on the country--fostering research and development, ensuring ethical AI use, preparing the workforce for AI-related jobs and prioritizing investments. Canada's national Artificial Intelligence Strategy was the first AI national strategy in the world with the aim of guiding AI policy priorities at a country level\footnote{\url{https://ised-isde.canada.ca/site/ai-strategy/en}. Accessed date: 26 November, 2023.}. Finland developed its national AI strategy also in 2017, closely followed by Japan, France, Germany and the United Kingdom in 2018. More than 30 other countries and regions have launched their national AI strategies as of 2021\footnote{\url{https://hai.stanford.edu/news/state-ai-10-charts}. Accessed date: 26 November, 2023.}, including Brazil.

    These national strategies and public policies are shaped and influenced somehow by global rankings of IA, in which countries are ranked by their AI capacity \cite{Erkkila2023}, in order to grasp the degree to which they are prepared to leverage its potential for reshaping business, government, and society\footnote{\url{https://www.tortoisemedia.com/wp-content/uploads/sites/3/2023/07/AI-Methodology-2306-4.pdf}. Accessed date: 26 November, 2023.}. So, given the positive and negative potential effects that the utilization of such AI metrics can have in the different society aspects, having robust indicators representing country's AI capabilities are essential for informed decision-making, resource allocation, and fostering AI innovation, thus contributing to economic growth, societal advancement, and global competitiveness.  

    In this paper, we intend to investigate the requirements that usually stands behind the development of robust indicators. For this purpose, we start from the indicator proposed by Tortoise Media\footnote{\url{https://www.tortoisemedia.com/wp-content/uploads/sites/3/2023/07/AI-Methodology-2306-4.pdf}}, The Global AI Index (GAII). GAII was the first index proposed to rank countries based on their AI capacity. This index is comprised by 7 dimensions (infrastructure, operating environment, research, development, commercial ventures, government strategy) that are grouped by associative themes around three main pillars: investment, innovation and implementation of AI, serving therefore as a composite index. To each AI dimension used in the index construction is given a relative importance weight, intending to consider the fact that contributions of different dimensions have different impact degrees into the AI capacity. In order to construct The Global AI Index, Tortoise adopted deterministic weights based on the comprehensiveness of the source dataset. By using the weighted sum aggregation procedure, the Global AI Index is calculated and then used to rank the 62 countries.

    From the Global AI Index description, our focus is to discuss such an index from a methodological perspective. From the ten-step process guide outlined in the handbook on composite indicators~\cite{joint2008handbook}, we aim to scrutinize three, which can enhance the transparency and quality of such as indicator. These characteristics include {Step 5}, which pertains to the weights employed; {Step 6}, addressing the aggregation of data; and {Step 8}, concerning the robustness and sensitivity analysis of the final ranking through weight adjustments. We assume that the other seven steps undertaken by Tortoise are appropriate, and therefore, their discussion is beyond the scope of our study.

    The three steps we examine here are of a particular interest. As highlighted by \cite{Greco_2019}, indicators considering multiple dimensions should undergo careful aggregation and weighting by addressing correlation and compensability issues among dimensions and avoiding subjective weighting approaches. Numerous studies, including those by~\cite{becker2017weights, pinar2022choquet}, underscore the critical role of weights and aggregation techniques in constructing composite indicators. The European Commission, in its communication on better regulation\footnote{\url{https://commission.europa.eu/law/law-making-process/planning-and-proposing-law/better-regulation/better-regulation-guidelines-and-toolbox_en}}, also delves into the sensitive nature of employing weights. They advocate for justifying weight choices based on robust ethical, scientific, or legal arguments~\cite{europeancomission}.  Besides, the Commission recommends conducting sensitivity and robustness analyses to enhance transparency regarding the assumptions underlying weight values. In GAII, the problem subjectivity in the definition of weights was even noted by Tortoise itself \cite{Tortoise}.

    In order to address these three steps in the construction of IA index, our goal is to answer the following three research questions:
    
    \begin{enumerate}
        \item Is the hypothesis of interactions between AI dimensions true? If so, how should these interactions be considered in order to mitigate bias in the resulting AI ranking?
        \item Do the criteria weights influence the resulting AI ranking of countries?
        \item Is it possible to determine a robust AI ranking without defining the weights in a deterministic manner?
    \end{enumerate}

 To answer these questions, we employ advanced techniques of Multiple Criteria Decision Analysis (MCDA)~\citep{roy1985methodologie}. This approach, renowned for its ability to address the inherent complexity of considering multiple dimensions, offers a robust and nuanced perspective on indicator composite development\cite{Greco_2019, munda2020use, europeancomission}. To explore the first question, we utilize the Pearson correlation coefficient~\cite{pearson1900x} to evaluate the interactions between criteria and demonstrate that there are redundancy among some of them. Thereafter, we apply a non-linear aggregation operator known as the Choquet integral~\cite{choquet1954theory}. As in the Choquet integral one is able to model intercriteria interactions, such as complementary and/or redundancy effects, the use of supplementary information about the data can help to prevent biased outcomes. We then compare the rankings obtained by the Choquet integral and the GAII, whose weights are the same as proposed in Tortoise.
    
    To address the second question, we employ the MCDA method known as Stochastic Multicriteria Acceptability Analysis (SMAA). SMAA is a widely used approach for assessing the robustness of weighting mechanisms in decision models. It has been designed for modeling decisions in situations where certain parameters, such as the weights, are either unknown or challenging to be predefined. These uncertain parameters are addressed through Monte Carlo simulation. Among the array of descriptive measures provided by SMAA, one of the key outputs is the probability of each country's ranking position. Consequently, by means of the utilization of SMAA, we can analyze the robustness of the proposed GAII ranking when weights are altered.
  
    Finally, to address the third question, we derive rankings based on the Condorcet method~\cite{de2014essai, Brandt2016} after applying the SMAA technique using both the weighted sum and the Choquet integral. The proposed rankings achieve a more robust result since the weights are not deterministic and the procedure is based on a well-established approach to rank candidates given preference information over them.

    In a previous study \cite{BRACIS2023}, we presented a preliminary critical analysis of 2022 GAII. However, in this current paper, we combine the SMAA methodology and the Choquet integral in order to conduct a robustness analysis with respect to model structure. Moreover, we also apply the Condorcet method to the results provided by SMAA aiming at proposing a new AI ranking of the countries. It is worth mentioning that, in 2023, Tortoise released an updated GAII which we are taking into consideration in this study.

    This paper is organized as follows. In Section~\ref{sec:mcda}, we introduce the Multiple Criteria Decision Analysis problem and provide the theoretical background on the SMAA methodology and the Choquet integral. Section~\ref{sec:tortoise} presents an overview of the methodology used in Tortoise to derive the Global AI Index. Section \ref{sec:proposal} presents the methodology adopted in this paper in order to answer the three research questions set above. Results are presented and discussed in Section \ref{sec:results}. We conclude this paper in Section~\ref{sec:conclusions}. 


\section{Multiple Criteria Decision Analysis}
\label{sec:mcda}
    Multiple Criteria Decision Analysis is a research field focused on mathematical and computational design tools that can be employed by either ranking or classifying alternatives. One of the key advantages of MCDA is its ability to consider multiple perspectives, preferences, and uncertainties. Through techniques such as Analytic Hierarchy Process (AHP), Analytic Network Process (ANP), and outranking methods like PROMETHEE, MCDA allows for the incorporation of diverse stakeholder opinions and the assessment of trade-offs between criteria. These methods enable the construction of more robust and context-specific indicator systems \cite{Greco_2019, munda2020use, europeancomission, campello2023exploiting}.
    
    In MCDA, alternatives $\mathcal{A} = \{a_{1}, a_2, \ldots, a_m\}$ are evaluated according to a set of criteria $\mathcal{G} = \{g_1, g_2, \ldots, g_n\}$ which represent relevant attributes to be considered in the decision process. One generally represents by $\mathbf{X}_{m \times n}$ the decision matrix, whose each row $i$, $i=1, \ldots, m$, indicates the criteria values for alternative $a_i$. Each element of $\mathbf{X}_{m \times n}$ is denoted by $g_j(a_i)$, which indicates the performance of the alternative $a_i$ with respect to the criterion $g_j$, $j=1, \ldots, n$. In the case of the weighted sum aggregation, associated to each criterion $g_j$, one defines a relative importance $w_j$ called criterion weight. We denote by $\mathbf{w} = (w_1, w_2, \ldots, w_n)$ the (criteria) weight vector. A crucial step in many MCDA techniques is the aggregation procedure, which maps each row $i$ of the decision matrix into a scalar that can be used as a score for alternative $i$. In the ranking context, one can determine a ranking for the alternatives by sorting their scores. 

    
\subsection{Choquet integral}
\label{ssec:agregacao}
 
The weighted sum method is a widely employed aggregation method in many real-world decisions, including in the construction of GAII (Equation~\eqref{eq_WS}), mainly due its simplicity. This method may be suitable when statistical dependence among criteria holds. However, in cases with interactions among criteria, applying the weighted sum method, overlooking this structural characteristic in the data, may introduce bias in the obtained ranking. For instance, if two criteria have a high positive correlation, aggregating then involves counting the same latent information twice. In such scenarios, the Choquet integral~\cite{choquet1954theory} emerges as a preferable option~\cite{Grabisch1996,Grabisch2010,Grabisch2016}. By using the Choquet integral, one is able to model negative (or positive) interactions among criteria, which is useful, for instance, to deal with correlations in the decision data~\cite{Marichal2000}. Therefore, one can soften the impact of correlated criteria in the aggregation procedure, which contributes to enhance the robustness of obtained ranking.

    The discrete Choquet integral (CI) is a non-linear aggregation function defined as follows:
	\begin{equation} \label{eq:eq_choquet}
            s_i^{CI} = \sum_{j=1}^{n}\left[g_{(j)}(a_i) - g_{(j-1)}(a_i)\right] \mu \left(\left\{(j), \ldots, (n) \right\} \right),
        \end{equation}
where $(1), \ldots, (n)$ indicates a permutation of indices $j$ such that $0 = g_{(0)}(a_i) \leq g_{(1)}(a_i) \leq g_{(j)}(a_i) \leq \ldots \leq g_{(n)}(a_i)$ and $\mu(\cdot)$ represents the set of parameters known as capacity coefficients. A capacity $\mu:2^{N} \rightarrow \mathbb{R}_{+}$ is a set function defined on the power set of $N=\left\{1, 2, \ldots, n\right\}$ (the set of criteria) satisfying the following axioms:
        \begin{itemize}
             \item [(a)] Normalization: $\mu(\emptyset) = 0$ and $\mu(N) = 1$,
             \item [(b)] Monotonicity: $\forall A \subseteq B \subseteq N, \mu(A) \leq \mu(B)$.
        \end{itemize}

An interesting property of the Choquet integral parameters is that they can be represented by means of the Shapley values and Shapley interaction indices~\cite{Grabisch1996,Grabisch1997alt}. The Shapley values~\cite{shapley1953value} is a well-established concept in cooperative game theory which provides a fair division of the total game payoff among its players. In the context of the MCDA, the Shapley values will then indicate the marginal contribution of each criterion into the aggregation procedure. The Shapley value of a criterion $g_j$ is defined by

\begin{equation}
\label{eq:power_ind_s}
\phi_{j} = \sum_{A \subseteq N\backslash j} \frac{\left(n-\left|A\right|-1\right)!\left|A\right|!}{n!} \left[\mu(A \cup \left\{j\right\}) - \mu(A) \right],
\end{equation}
where $\left| A \right|$ indicates the cardinality of the subset $A$, $\phi_j \geq 0$ and $\sum_{j=1}^{n} \phi_j = 1$. By making a parallel between the Shapley value and the weights in the weighted sum, both $\phi_j$ and $w_j$ can be interpreted as the relative importance of the associated criterion into the aggregation function.
    
    In addition to assessing marginal contributions, one can also analyze the interaction between pairs of criteria\footnote{We could also evaluate the interaction for coalitions of more than two criteria. However, as the interpretation for such coalitions are not straightforward, we only consider interactions between pairs of criteria in our analysis.}. The Shapley interaction index between two criteria $\{g_j, g_{j'}\}$ is defined as follows:
		
\begin{equation}
\label{eq:int_pair_ind_s}
I{j,j'} = \sum_{A \subseteq N\backslash \left\{j,j'\right\}} \frac{\left(n-\left|A\right|-2\right)!\left|A\right|!}{\left(n-1\right)!} \left[\mu(A \cup \left\{j,j'\right\}) - \mu(A \cup \left\{j\right\}) - \mu(A \cup \left\{j'\right\}) + \mu(A)\right],
\end{equation}
where $I_{j,j'} \in [-1,1]$ and can be interpreted as the interaction degree of coalition of criteria $i,i'$. These indices offer a direct and practical interpretation about the degree of interaction between criteria: $I_{j, j'} > 0$ models synergy between $\{g_j, g_{j'}\}$ (or complementary effect), $I_{j, j'} < 0$ represents redundancy between $\{g_j, g_{j'}\}$ (or negative effect), and $I_{j, j'} = 0$ indicates no interaction between $\{g_j, g_{j'}\}$ (criteria act independently). 

A drawback in the Choquet integral that can be observed from Equation~\eqref{eq:eq_choquet} is that the number of parameters to be elicited may be large. Indeed, it exponentially increases with the number of criteria. With the aim of reducing the number of parameters to be elicited, we adopt in this paper the 2-additive capacity representation introduced by~\citep{grabisch1997k}. The presence of a 2-additive capacity implies that interactions only exist between pairs of criteria, and we assume that the interaction among three or more criteria is negligible. In this case, the Choquet integral can be defined by mans of the Shapley values and the interaction indices as follows:

\begin{eqnarray}
            s_i^{2adCI} &=& \sum_{I_{j,j'} > 0} \min \{g_j(a_i), g_{j'}(a_i)\}I_{j,j'} + \sum_{I_{j,j'} < 0} \max \{g_j(a_i), g_{j'}(a_i)\}|I_{j,j'}| \nonumber\\&+& \sum_{j=1}^{n} g_{j}(a_i) (\phi_{j} - \frac{1}{2}\sum_{j'\neq j}|I_{j,j'}|).\label{eq:choquet_shapley}
\end{eqnarray}

One may note from Equation~\eqref{eq:choquet_shapley} that the numbers of parameters drastically reduced from $2^n$ to $2^n$. As in this equivalent formulation of the Choquet integral one also needs to ensure the axioms of a capacity, one must satisfy the following conditions (see~\cite{grabisch2000graphical} for further details):

    \begin{equation}\label{eq_cond_index_bound}
            \sum_{j=1}^{n} \phi_j = 1
    \end{equation}
    and
    \begin{equation}\label{eq_cond_index_monot}
            \phi_{j} - \frac{1}{2}\sum_{j' \neq j}|I_{j,j'}| \geq 0, \forall j \in N.
    \end{equation}
    
    Since it is challenging for the decision-maker to determine the Shapley interaction indices, an alternative is to employ an unsupervised approach to learn suitable values based on the correlation coefficient extracted from data~\cite{duarte2017novel}. In this study, we utilize two distinct approaches to estimate the Shapley index. The first approach, proposed by~\cite{duarte2017novel} (see~\cite{Pelegrina2020} for an alternative formulation which simplifies the optimization problem), is referred to as Unsupervised 1 (u1) and is detailed in Subsection~\ref{sssec:unsupervised_approach1}. The second approach, introduced by~\cite{Pelegrina2024}, is designated as Unsupervised 2 (u2) and is outlined in Subsection~\ref{sssec:unsupervised_approach2}. The u2 technique represents an improvement over u1, as it reduces bias in determining the Shapley interaction indices. We assess the performance of both methods on our dataset, testing the effectiveness and draw conclusions about their suitability for our application.
    
    \subsubsection{Unsupervised approach 1 to learn the Choquet integral parameters (u1)}\label{sssec:unsupervised_approach1}             

    The Choquet integral-based unsupervised approach is used here to automatically adjust the Shapley interaction indices. The goal is to achieve interaction indices $I_{j,j'}$ as close as possible from the negative of a similarity measure $\rho_{j,j'}$ between pairs of criteria, such as the Pearson correlation coefficient between them~\cite{pearson1900x}. For instance, if two criteria $g_j, g_{j'}$ are positively correlated ($\rho_{j,j'} > 0$), one should assign a negative interaction index for them ($I_{j,j'} < 0$) in order to model a redundant effect. In this formulation, the optimization problem used to automatically adjust the Shapley interaction indices is given as follows~\cite{duarte2017novel,Pelegrina2020}:
        \begin{equation}
        \label{eq:opt_prob_nonsuperv}
        \begin{array}{ll}
        \displaystyle\min_{I_{j,j'}, \forall j,j' \in N} & \sum_{j,j'}\left(I_{j,j'} + \rho_{j,j'}\right)^2 \\
        \text{s.t.} & \phi_{j} - \frac{1}{2} \sum_{j' \neq j} \pm I_{j,j'} \geq 0, \, \, \forall j \in N \\
         & \sum_{j} \phi_{j} = 1
        \end{array},
        \end{equation}
where $\pm$ in the first constraint avoids the use of absolute values in $I_{j,j'}$. Note that, in this optimization problem, we do not find the Shapley values, they should be predefined. Moreover, as it is a quadratic problem, it can be easily tackled by most of the available solvers.
    
    \subsubsection{Unsupervised approach 2 to learn the Choquet integral parameters (u2)}
    \label{sssec:unsupervised_approach2}

    The unsupervised approach 1 is a suitable method for learning the Choquet integral Shapley parameter. However, as discussed in~\cite{Pelegrina2024}, u1 may exhibit bias, some $I_{j,j'}$ might be much closer to $\rho_{j,j'}$ than others parameters $I_{j'',j'''}$ to $\rho_{j'',j'''}$. Aiming at overcoming this inconvenience,~\cite{Pelegrina2024} proposed a novel Choquet integral-based unsupervised approach to automatically defined the Shapley interaction indices. The key idea consists in minimizing the difference between all $I_{j,j'}$ and $-\rho_{j,j'}$ while ensuring consistent ratio $t$ between them. In other words, we ensure that $t = -I_{j,j'}/\rho_{j,j'}$ for all $j,j'$. This condition is satisfied by solving the following optimization problem:      
        \begin{equation}
        \label{eq:opt_prob_nonsuperv}
        \begin{array}{ll}
        \displaystyle\max_{t, \textbf{I}} & t \\
        \text{s.t.} & \phi_{j} - \frac{1}{2} \sum_{j' \neq j} \mbox{sign}(-\rho_{j,j'}) I_{j,j'} \geq 0, \, \, \forall j \in N \\
        & I_{j,j'} - \mbox{sign}(-\rho_{j,j'}) t \rho_{j,j'} = 0, \, \, \forall j,j' \in N \\
         & \sum_{j} \phi_{j} = 1\\
         & 0 \leq t \leq 1
        \end{array},
        \end{equation}
where sign(z) is the sign function: $\text{sign}(z) = 1$ if $z \geq 0$ and -1 otherwise. The ratio $t$ is confined to the range $[0,1]$. One may note that the first and third constraints mirror those of the u1 approach, while the newly introduced constraint aims to guarantee the consistency of the ratio for all pairs of criteria.

    \subsection{Weighting robustness analyses}
    \label{ssec:smaa}
   
    SMAA stands out as a simulation-driven technique applicable to discrete multicriteria decision-making issues, especially in scenarios where model parameters, including criteria weights, exhibit uncertainty, imprecision, or are partially or wholly unavailable \cite{Lahdelma1998, Pelissari_review}.  Uncertain criteria information are represented by stochastic variables with joint density function. During a Monte Carlo simulation procedure, uncertain variable values are drawn from their respective distributions, and alternatives are assessed using the decision model, which may involve a weighted average or any other aggregation method. In the context of ranking problems, SMAA identifies all potential rankings for the alternatives, expressing the outcomes in probabilistic terms. Thus, instead of providing a deterministic rank, SMAA generates a matrix where the rows represent the alternatives and the columns denote their potential positions in the ranking. Each element in this matrix is known as \textit{rank acceptability index}, $b_{i}^s$, and represents the probability of alternative $i$ being positioned at $s$. This index ranges between $0$ and $1$, and the closer $b_{i}^s$ is to 1, the greater the probability of $a_{i}$ being in position $s$. The SMAA method can also be applied in the sorting context (\cite{Tervonen2009a, Lahdelma2010, PELISSARI2019235, arcidiacono2020robust, PELISSARI2021102381,  PELISSARI2022108727, PELISSARI2022117898}).

   Various descriptive measures for alternatives have been proposed in the literature \cite{Pelissari_review}. One such measure is the central weight vector $w_{i}^c$, which enables the decision-maker to learn the weights that result in an alternative being ranked first, rather than predefining them before addressing the problem. Additionally, the confidence factor $p^{c}_i$ represents the probability of an alternative being the preferred one, given the criteria weights expressed by its central weight vector. Moreover, the pairwise winning index $c_{ik}$ can be utilized to quantify the extent to which one alternative outperforms another, considering the stochastic weight distribution. In essence, the pairwise winning index represents the probability of one alternative being better to another. The pairwise winning indices can also be employed to create a probabilistic ranking of the alternatives, as elaborated in Section~\ref{ssec:condorcet}.

    SMAA can be applied with a complete lack of information, where weights are generated following a uniform distribution. This technique can also be used when partial preference information is available, such as specifying the ordinal relevance of the criteria~\cite{menou2022multicriteria}.

 \subsection{Obtaining a robust ranking}
\label{ssec:condorcet}

In order to derive a ranking from the probabilities given by SMAA, 
we can employ an approach based on the Condorcet method. The Condorcet method, as described in~\cite{de2014essai, Brandt2016}, is a preferential voting system designed to determine the candidate who would prevail in a pairwise comparison against every other candidate. In this method, each voter submits ranked preferences for the candidates. Pairwise comparisons are then conducted between each possible pair of candidates, evaluating which candidate is preferred over the other. The candidate who consistently emerges as the preferred choice in every individual comparison is deemed the Condorcet winner. If a clear Condorcet winner exists, that candidate is declared the overall winner. However, in cases where a circular tie occurs, known as the \textit{Condorcet paradox} or \textit{cycle}, with each candidate defeating another but being defeated by a different candidate, additional methods, such as those outlined in~\cite{nurmi2012comparing, gehrlein1997condorcet}, may be necessary to resolve the deadlock.

  This approach can be applied within the context of MCDA to derive a ranking through pairwise comparisons in the SMAA method. Thus, for a pair of alternatives, $a_i$ and $a_k$, it is possible to compute a pairwise winning index, $c_{ik}$, representing the probability of alternative $i$ being more preferred than $k$. The procedure is as follows. In each SMAA simulation, if the score of $a_i$ is greater than the score of $a_k$, the counter $C_{ik}$ is incremented. Conversely, if the score of $a_k$ is greater than the score of $a_i$, the counter $C_{ki}$ is incremented. It is noteworthy that, with real-valued functions and continuous distributions for criteria weights, occurrences of identical scores for two alternatives are rare~\cite{montes2020correspondence, menou2022multicriteria}. At the end of the $S$ simulations, the pairwise winning indices are computed as:
 
 $$c_{ik} = C_{ik}/S, $$
 
 \noindent for all pairs of alternatives, where $c_{ik} \in [0,1]$ and $c_{ik} + c_{ki} = 1$. If $c_{ik} = 1$, it indicates total preference for $a_i$ over $a_k$, and if $c_{ik} > 0.5$, it means that $a_i$ is (\textit{statistically}) preferred to $a_k$~\cite{montes2020correspondence, menou2022multicriteria}. Thus, the complete ranking is obtained by performing pairwise comparisons of the winning indices.

 However, the \textit{Condorcet paradox} can occur among three alternatives, $a_i$, $a_k$, and $a_\ell$, if $c_{ik} = c_{k\ell} = c_{\ell i} = 2/3$, and $c_{ki} = c_{\ell k} = c_{i \ell} = 1/3$. In this study, we use the Schulze method~\cite{schulze2018schulze} to resolve the cycle.


    \section{The Global AI Index: Tortoise methodology}
\label{sec:tortoise}

The Global AI Index (GAII) aims to assess the national ecosystems influencing AI capacity across 62 countries, denoted by [$a_1$, $a_2$, $\cdots$, $a_{62}$]. To determine the ranking, data were selected based on relevance, up-to-date sources, and alignment with key AI sector issues. These data are then categorized into three primary pillars: Implementation, Innovation, and Investment, further subdivided into seven sub-pillars or criteria. The Implementation pillar evaluates the operationalization of AI by practitioners, covering 3 criteria: talent, infrastructure, and the operating environment. The Innovation pillar scrutinizes technology breakthroughs and methodological advancements indicative of future AI capacity, encompassing 2 criteria: research and development. The Investment pillar assesses financial and procedural commitments to AI, including 2 criteria: commercial ventures and government strategy.

The data used to build GAII  comes from 28 different sources, encompassing government reports, public databases from international organizations, think tanks, private companies, and Tortoise's proprietary research. 

Country scores ($s_i$) are obtained through the weighted sum aggregation method, calculated as follows:
\begin{equation}\label{eq_WS}
    \text{s}_i = \sum_{j=1}^{n} w_j g_j(a_i), \forall i,
\end{equation}
where $g_j(a_i)$ represents the evaluation of criterion $j$, $j = 1, \cdots, 7$, for country $a_i$, $i = 1, \cdots, 62$, and $w_j$ denotes the weights assigned to the corresponding criterion. In the 2023 ranking, the Tortoise-chosen criteria (and weights) were as follows: Infrastructure ($w_1 = 0.11$), Operating Environment ($w_2 = 0.06$), Talent ($w_3 = 0.15$),  Development ($w_4 = 0.14$), Research ($w_5 = 0.26$), Commercial ventures ($w_6 = 0.24$), Government Strategy ($w_7 = 0.04$). All data, final ranking, and Methodology Report used by Tortoise can be found in~\cite{Tortoise}.
  

 \section{Methodology}
  \label{sec:proposal}

 To address the three research questions set in this study, we propose the following methodologies (one for each research question).

 \begin{description}
     \item [Methodology 1:] This methodology was proposed to address the first question concerning the interaction between criteria. To tackle this issue, we assumed as the Shapley values ($\phi_j$) the deterministic weights ($w_j$), as proposed by Tortoise and detailed in Section~\ref{sec:tortoise}. Subsequently, we calculate the correlations between pairs of criteria using Pearson correlation coefficients ($\rho_{j,j'}$, where $j,j' = 1, \ldots, 7$). We then apply the unsupervised approaches described in Subsections~\ref{sssec:unsupervised_approach1} and~\ref{sssec:unsupervised_approach2} to derive the interaction indices $I_{j,j'}^{u1}$ and $I_{j,j'}^{u2}$, respectively. Finally, we apply Equation~(\ref{eq:choquet_shapley}) using $I_{j,j'}^{u1}$ and $I_{j,j'}^{u2}$ to obtain the rankings denoted as $CI^{u1}$ and $CI^{u2}$, respectively.

     \item [Methodology 2:] The aim here is to analyze the second question concerning the influence of criteria weights on the final ranking. To achieve this, we utilize the SMAA technique, where weights are randomly generated. In the aggregation step, we apply both the weighted sum and Choquet integral (employing both the u1 and u2 unsupervised learning methods). The outputs are the rank acceptability indices for each of the three approaches (weighted sum, Choquet + u1, and Choquet + u2).

     \item  [Methodology 3:] This approach was developed to address the third question, aiming to obtain a robust AI ranking without relying on deterministic weights. To achieve this, the rank acceptability index obtained using both the weighted sum and Choquet integral (utilizing both the u1 and u2 unsupervised learning methods) is subjected to the Condorcet method, as explained in Subsection~\ref{ssec:condorcet}. This process yields three final rankings: $WS$-Cond., $CI^{u1}$-Cond., and $CI^{u2}$-Cond.
   
 \end{description}

 \section{Results and Discussion}
\label{sec:results}

 As discussed earlier, GAII was constructed using the weighted sum method and deterministic weights proposed by Tortoise. As highlighted by Tortoise itself, the selected weights are grounded in subjective assumptions, with one of these assumptions concerning the completeness of the data \cite{Tortoise}. This subjective decision has the potential to influence the scoring of countries and subsequently impact their positions in the rankings.

In this section, we conducted analyses on the 2023 GAII proposed by Tortoise, utilizing the same data (available on their \href{https://www.tortoisemedia.com/intelligence/global-ai/#rankings}{website}) for the 62 countries as alternatives and applying the identical set of seven AI criteria: Infrastructure ($g_1$), Operating Environment ($g_2$), Talent ($g_3$), Development ($g_4$), Research ($g_5$), Commercial ventures ($g_6$), Government strategy ($g_7$). The Python code used to obtain the results can be found at \href{https://github.com/BSCCampello/aidevelopmentindices}{this link}.

 In a first analysis, presented in Subsection~\ref{ssec:resultadoschoquet}, our goal is to verify the hypothesis of interactions between AI dimensions and investigate if there are differences in the final ranking when considering interactions between criteria using the same weights as in GAII. To achieve this, we use Methodology 1 to obtain $CI^{u1}$ and $CI^{u2}$ rankings.
 
 In a second analysis, presented in Subsection~\ref{ssec:resultados_pesos_aleatorios}, the objective is to perform a robustness analysis for weighting by applying SMAA, assuming a complete absence of weight information. Methodology 2 is employed to analyze sensitivity to changes in weights, while Methodology 3 is utilized to compare the robustness between the GAII ranking and our proposed rankings, namely $WS$-Cond., $CI^{u1}$-Cond., and $CI^{u2}$-Cond.

 The third analysis, outlined in Subsection~\ref{ssec:resultados_pesos_ordenado}, is akin to the second one, but in SMAA, weights adhere to an ordinal preference, following the same preference order as used to derive the GAII ranking: Research $\succ$ Commercial Venture $\succ$ Talent $\succ$ Development $\succ$ Infrastructure $\succ$ Operating Environment $\succ$ Government Strategy.  Thus, similar to the second analysis, Methodologies 2 and 3 are employed to analyze sensitivity to changes in weights and to compare the robustness between the GAII ranking and our proposed rankings, $WS$-Cond., $CI^{u1}$-Cond., and $CI^{u2}$-Cond.

 To measure the difference between the countries ranking in GAII and in our approaches, we use the Kendall's tau distance~\cite{kendall1938new}. This coefficient quantifies the degree of similarity between two rankings when comparing the same set of objects. The Kendall tau value represents the number of pairwise inversions required to align both rankings~\cite{abdi2007kendall}. For two rankings, denoted as \textbf{r} and $\hat{\textbf{r}}$, the Kendall tau distance between them can be expressed as $\left\| \textbf{r} - \hat{ \textbf{r}} \right\|_\kappa$ or $\tau_{\textbf{r}\times\hat{\textbf{r}}}$, and it is computed as follows:

\small
\begin{align}
	\tau_{\textbf{r}\times\hat{\textbf{r}}} = \dfrac{2}{n(n-1)} | \{(q, u): q<u, (\tau_\textbf{r}(q) < \tau_\textbf{r}(u)\ \land  \tau_{\hat{\textbf{r}}}(q) > \tau_{\hat{\textbf{r}}}(u))\nonumber\\   
	\lor    (\tau_\textbf{r}(q) > \tau_\textbf{r}(u)\ \land  \tau_{\hat{\textbf{r}}}(q) < \tau_{\hat{\textbf{r}}}(u))\}  |,
\end{align}
\normalsize

\noindent where $\tau_\textbf{r}(q)$ and $ \tau_{\hat{\textbf{r}}}(q) $ represent the positions of element $q$ in vectors $\textbf{r}$ and $ \hat{\textbf{r}} $, respectively. The $\tau_{\textbf{r}\times\hat{\textbf{r}}}$ value is constrained between zero and one; a value of zero indicates identical rankings, while a value of one signifies opposite rankings.

\subsection{Exploring interactions among criteria considering the same weights as proposed in Tortoise}
\label{ssec:resultadoschoquet}

Firstly, we verified that there are interactions between the AI dimensions by calculating the correlation coefficients (a described in the second column of Table~\ref{tab:interactions}).  Then, in order to investigate the impact of interactions between criteria in the final ranking, we apply both u2 and u1 unsupervised approaches and obtained the interaction indices presented in the third and fourth columns of Table~\ref{tab:interactions}, respectively. We highlight rows with significant correlations (e.g., $\rho_{j,j'} \geq 0.70$), which indicates redundancies within the dataset (see, for instance, the correlations between criteria $\{g_3, g_{5}\}$, $\{g_4, g_{5}\}$, $\{g_4, g_{6}\}$and $\{g_5, g_{6}\}$). 

 \begin{table}[ht]
\centering
\small
\caption{Correlation coefficients and interaction indices.}
\begin{tabular}{c|c|c|c}
\textbf{\begin{tabular}[c]{@{}c@{}}Coalition of\\ criteria j,j'\end{tabular}} & \textbf{\begin{tabular}[c]{@{}c@{}}Correlation coefficient\\ ($\rho_{j,j'}$)\end{tabular}} & \textbf{\begin{tabular}[c]{@{}c@{}}Interaction index\\ ($I_{j,j'}^{u2}$)\end{tabular}} & \textbf{\begin{tabular}[c]{@{}c@{}}Interaction index\\ ($I_{j,j'}^{u1}$)\end{tabular}} \\ \hline
1,2                                                                           & 0.5435                                                                                     & -0.0181                                                                                 & -0.1008                                                                                 \\ \hline
1,3                                                                           & 0.3607                                                                                     & -0.0120                                                                                 & 0.0000                                                                                  \\ \hline
1,4                                                                           & 0.5868                                                                                     & -0.0196                                                                                 & 0.0000                                                                                  \\ \hline
1,5                                                                           & 0.6186                                                                                     & -0.0206                                                                                 & -0.0506                                                                                 \\ \hline
1,6                                                                           & 0.5226                                                                                     & -0.0174                                                                                 & -0.0078                                                                                 \\ \hline
1,7                                                                           & 0.5338                                                                                     & -0.0178                                                                                 & -0.0608                                                                                 \\ \hline
2,3                                                                           & 0.3503                                                                                     & -0.0117                                                                                 & 0.0000                                                                                  \\ \hline
2,4                                                                           & 0.3320                                                                                     & -0.0111                                                                                 & 0.0000                                                                                  \\ \hline
2,5                                                                           & 0.3332                                                                                     & -0.0111                                                                                 & 0.0000                                                                                  \\ \hline
2,6                                                                           & 0.2464                                                                                     & -0.0082                                                                                 & 0.0000                                                                                  \\ \hline
2,7                                                                           & 0.5136                                                                                     & -0.0171                                                                                 & -0.0192                                                                                 \\ \hline
3,4                                                                           & 0.6338                                                                                     & -0.0212                                                                                 & -0.0136                                                                                 \\ \hline
\rowcolor{Gray}
3,5                                                                           & 0.7565                                                                                     & -0.0253                                                                                 & -0.1572                                                                                 \\ \hline
3,6                                                                           & 0.6754                                                                                     & -0.0225                                                                                 & -0.1293                                                                                 \\ \hline
3,7                                                                           & 0.3913                                                                                     & -0.0131                                                                                 & 0.0000                                                                                  \\ \hline 
\rowcolor{Gray}
4,5                                                                           & 0.8534                                                                                     & -0.0285                                                                                 & -0.1179                                                                                 \\ \hline
\rowcolor{Gray}
4,6                                                                           & 0.8309                                                                                     & -0.0277                                                                                 & -0.1485                                                                                 \\ \hline
4,7                                                                           & 0.3483                                                                                     & -0.0116                                                                                 & 0.0000                                                                                  \\ \hline
\rowcolor{Gray}
5,6                                                                           & 0.8558                                                                                     & -0.0286                                                                                 & -0.1943                                                                                 \\ \hline
5,7                                                                           & 0.3815                                                                                     & -0.0127                                                                                 & 0.0000                                                                                  \\ \hline
6,7                                                                           & 0.2282                                                                                     & -0.0076                                                                                 & 0.0000                                                                                 
\end{tabular}
\label{tab:interactions}
\end{table}
 
 An important observation is that $g_5$ exhibits a significant correlation with other criteria, suggesting redundancy in its data concerning other criteria. As a result, it should be penalized in its weight to avoid bias. However, in deriving the GAII ranking, where correlation was not taken into account, $g_5$ received the highest weight value ($w_5 = 0.26$). These initial findings underscore the importance of analyzing criteria correlations for constructing an index and employing an appropriate method to address them. 

 A second remark in Table~\ref{tab:interactions} regards the difference of the interaction indices between approaches u2 and u1. One can observe that u2 distributes the values of $I_{j,j'}^{u2}$ more evenly based on the $\rho_{j,j'}$ values; specifically, higher the value of $\rho_{j,j'}$, lower the interaction degree $I_{j,j'}^{u2}$. In contrast, u1 yielded $I_{1,4}^{u1} = 0$ when $\rho_{1,4} = 0.5868$, and $I_{1,6}^{u1} = -0.0078$ when $\rho_{1,6} = 0.5226$. This implies that higher $\rho_{j,j'}$ values do not consistently translate to lower $I_{j,j'}^{u1}$, indicating a lack of equity in these values, potentially resulting in biased outcomes.
 
 Table~\ref{tab:ranking} displays the top ten countries in the GAII ranking, along with the rankings obtained using our proposed method, $CI^{u1}$ and $CI^{u2}$. We apply the Choquet integral with same weights as those used in GAII, and the interaction index detailed in Table~\ref{tab:interactions}. One can observe that the ranking obtained using $CI^{u2}$ is more similar to GAII than the ranking using $CI^{u1}$. However, notable changes include Canada and South Korea swapping positions at 5th and 6th in $CI^{u2}$ compared to GAII. Moreover, Israel and Germany exchanges positions at 7th and 8th. 

\begin{table}[ht]
\centering
\caption{Ranking from GAII and rankings using our proposal, applying the Choquet integral and utilizing the same criterion weights as in GAII, along with the interaction index shown in Table~\ref{tab:interactions}, for the top ten countries.}
\begin{tabular}{lccc}
\multicolumn{1}{c}{\multirow{2}{*}{\textbf{Position}}} & \multicolumn{3}{c}{\textbf{Countries}}  \\ \cline{2-4} 
\multicolumn{1}{c}{}                                   & GAII      & $CI^{u2}$  & $CI^{u1}$  \\ \hline
1st                                                    & USA         & USA         & USA         \\
2nd                                                    & China       & China       & China       \\
3rd                                                    & Singapore   & Singapore   & Singapore   \\
4th                                                    & UK          &  UK          & \cellcolor{Mint} South Korea \\
5th                                                    & \cellcolor{Peach} Canada      & \cellcolor{Mint} South Korea &  UK          \\
6th                                                    & \cellcolor{Mint} South Korea & \cellcolor{Peach} Canada      & \cellcolor{Butter} Germany     \\
7th                                                    &  \cellcolor{Lavender} Israel      & \cellcolor{Butter} Germany     & \cellcolor{Peach} Canada      \\
8th                                                    & \cellcolor{Butter} Germany     &  \cellcolor{Lavender} Israel      & Switzerland \\
9th                                                    & Switzerland & Switzerland &  \cellcolor{Lavender} Israel      \\
10th                                                   & Finland     & Finland     & India       \\ \hline
\end{tabular}
\label{tab:ranking}
\end{table}

  We calculate the Kendall tau distance to assess the difference between the GAII ranking and the ranking obtained using $CI^{u2}$, yielding $\tau_{GAII, CI^{u2}} = 0.2125$. Similarly, when comparing the GAII ranking with the one obtained using $CI^{u1}$, we observe  $\tau_{GAII, CI^{u1}} = 0.2866$. These results indicate that both rankings obtained using $CI^{u1}$ and $CI^{u2}$ differ from the GAII. Furthermore, they suggest that the ranking derived from $CI^{u2}$ is slightly closer to the GAII than the one obtained using $CI^{u1}$. In summary, these findings highlight variations in the ranking when evaluating the correlation between criteria using the same weight values as applied to derive the GAII.

\subsection{Robustness analysis of weights applying SMAA and Condorcet in the absence of weight information}
\label{ssec:resultados_pesos_aleatorios}

In this subsection, we assume a total absence of weight information, generating random weights following a uniform distribution, where $w_j \sim \mathcal{U}(0,1)$  $\forall j$. As a preliminary analysis, we constructed Table \ref{tab:ranking_condorcet_random}, which features the top ten countries in the GAII ranking alongside three rankings derived from our approach using Methodology 2 and 3. 

\begin{table}[ht]
\centering
\caption{Comparison of ranking positions for the top ten countries in GAII and those determined using our approaches, considering the absence of weight information.}
\begin{tabular}{lcccc}
\multicolumn{1}{c}{\multirow{2}{*}{\textbf{Position}}} & \multicolumn{4}{c}{\textbf{Countries}}  \\ \cline{2-5} 
\multicolumn{1}{c}{}                             & GAII       & $WS$ -  Cond.      & $CI^{u2}$ - Cond.   & $CI^{u1}$ -  Cond.   \\ \hline
1st                                              &   USA   & USA         & USA         & USA         \\
2nd                                              &  China    & China       & China       & China       \\
3rd                                              &  Singapore     & Singapore   & Singapore   & \cellcolor{Mint} South Korea \\
4th                                              &  \cellcolor{Lavender} UK    & \cellcolor{Mint} South Korea & \cellcolor{Mint} South Korea & Singapore   \\
5th                                              & \cellcolor{Peach} Canada     & \cellcolor{Butter} Germany     & \cellcolor{Butter} Germany     & \cellcolor{Butter} Germany     \\
6th                                             &  \cellcolor{Mint} South Korea    & \cellcolor{Peach} Canada      & \cellcolor{Peach} Canada      & \cellcolor{Peach} Canada      \\
7th                                              & Israel    & \cellcolor{Lavender} UK          & \cellcolor{Lavender} UK          & \cellcolor{Lavender} UK          \\
8th                                              & \cellcolor{Butter} Germany     & Finland     & Finland     & Finland     \\
9th                                              & Switzerland     & Japan       & Japan       & Japan       \\
10th                                             &  Finland    & Netherlands & Netherlands & Netherlands \\ \hline
\end{tabular}
\label{tab:ranking_condorcet_random}
\end{table}

\begin{figure}
    \centering
    \includegraphics[width=0.7\linewidth]{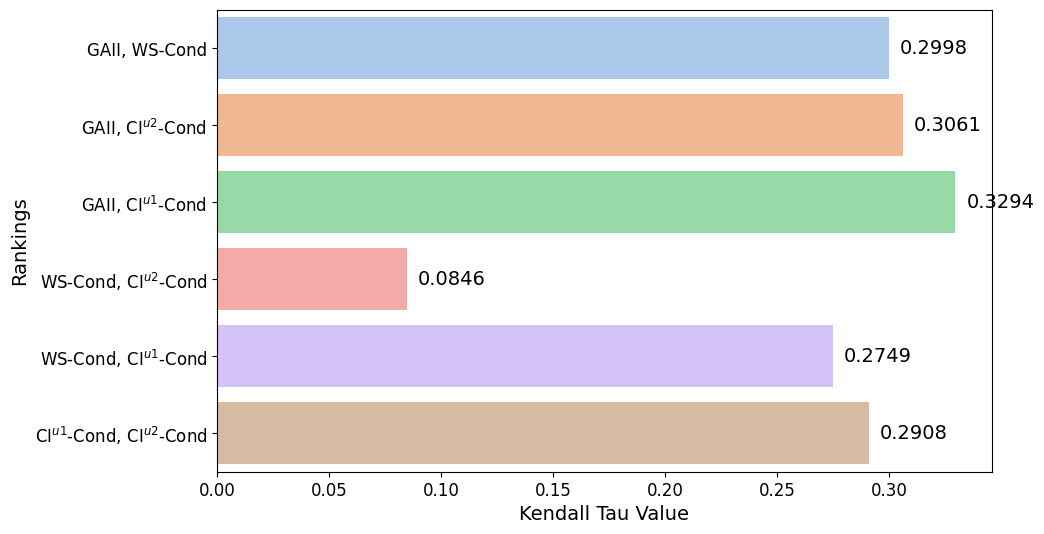}
    \caption{Kendall tau distances to measure differences between rankings.}
    \label{fig:tauentrerankings}
\end{figure}

Table \ref{tab:ranking_condorcet_random} shows that the USA, China, and Singapore secured the same positions in all rankings, except in the $CI^{u1}$ - Cond. ranking. This initial analysis may lead to the conclusion that the positions of the USA and China are robust since they do not depend on weights to occupy those positions. Furthermore, correlations between criteria do not affect their positions. In contrast, the UK, which holds the 4th position in GAII, is placed 7th following our approaches. A similar outcome is observed for Canada, which is ranked 5th in GAII but 6th in our approaches. 

To delve deeper into this analysis and identify disparities between these rankings, we created Figure~\ref{fig:tauentrerankings}, illustrating the Kendall tau distances obtained through pairwise comparisons of the complete rankings across all countries. From this figure, one can observe that the differences between GAII and our approaches are approximately 0.30, indicating significant distinctions. Hence, it demonstrates that a sensitivity analysis of the weights proves that distinct rankings can be achieved by varying weights.

While still analyzing the results shown in Figure~\ref{fig:tauentrerankings}, we observe that Rankings $WS$ - Cond. and $CI^{u2}$ - Cond. are very similar, whereas $WS$ - Cond. and $CI^{u1}$ - Cond. are significantly different. Since we found in Subsection~\ref{ssec:resultadoschoquet} that the $u2$ approach is fairer in penalizing bias in correlated criteria, we can deduce that, based on this data and with a total absence of information on weight values, considering the correlation between criteria does not lead to significantly different rankings.         

\begin{figure}[H]
\centering
\subfloat[$WS$ - GAII's ranking.]{\includegraphics[width=2.5in]{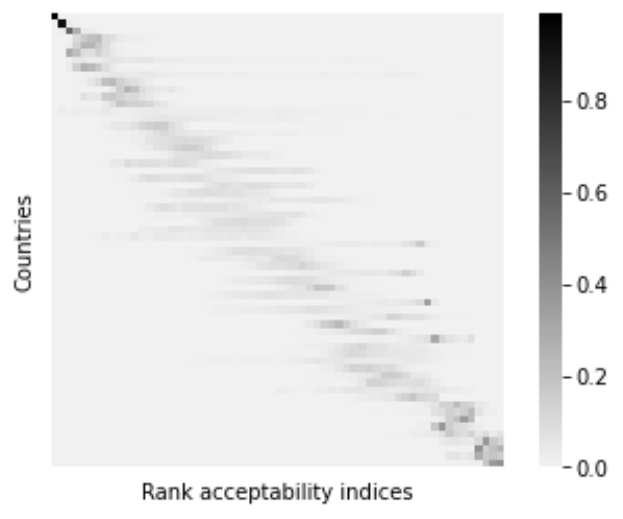}
\label{fig:smaa_ws}}
\hfill
\subfloat[$WS$ - Condorcet's ranking.]{\includegraphics[width=2.5in]{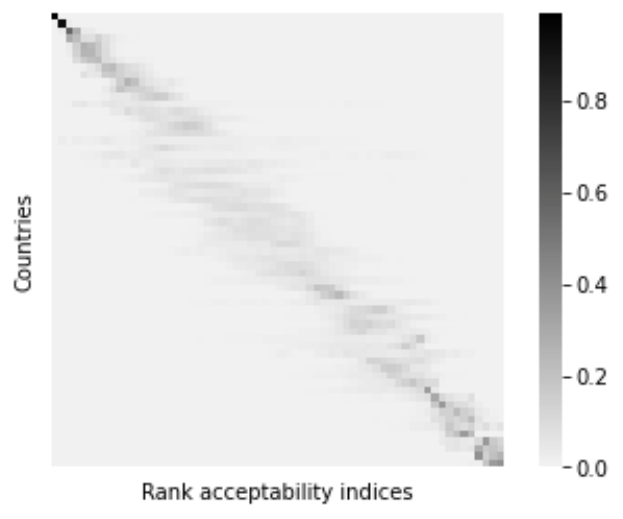}
\label{fig:smaa_ws_condorcet}}
\hfill
\subfloat[$CI^{u2}$ - GAII's ranking.]{\includegraphics[width=2.5in]{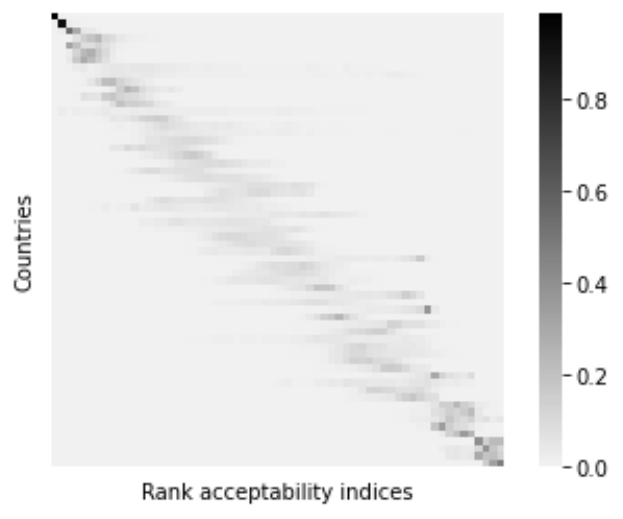}
\label{fig:smaa_ci}}
\hfill
\subfloat[$CI^{u2}$ - Condorcet's ranking.]{\includegraphics[width=2.5in]{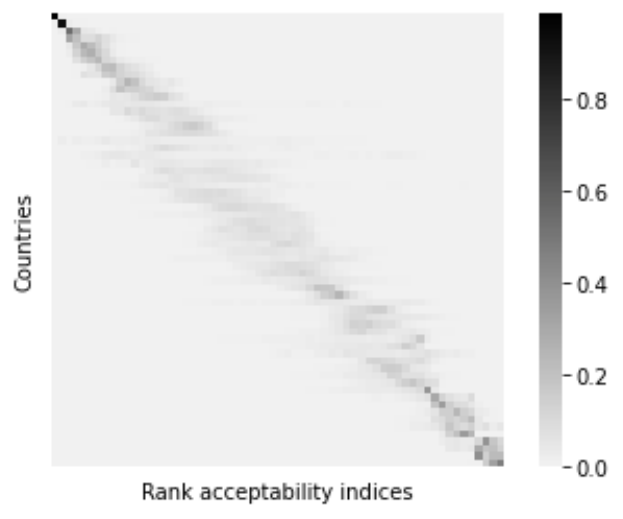}
\label{fig:smaa_ci_condorcet}}
\hfill
\subfloat[$CI^{u1}$ - GAII's ranking.]{\includegraphics[width=2.5in]{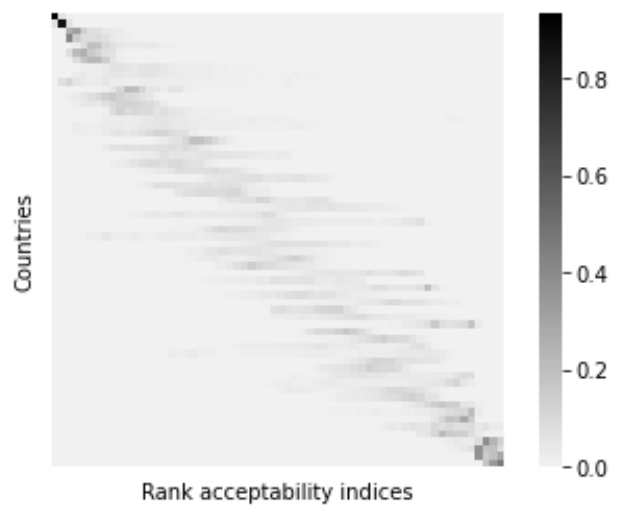}
\label{fig:smaa_ci_antigo}}
\hfill
\subfloat[$CI^{u1}$ - Condorcet's ranking.]{\includegraphics[width=2.5in]{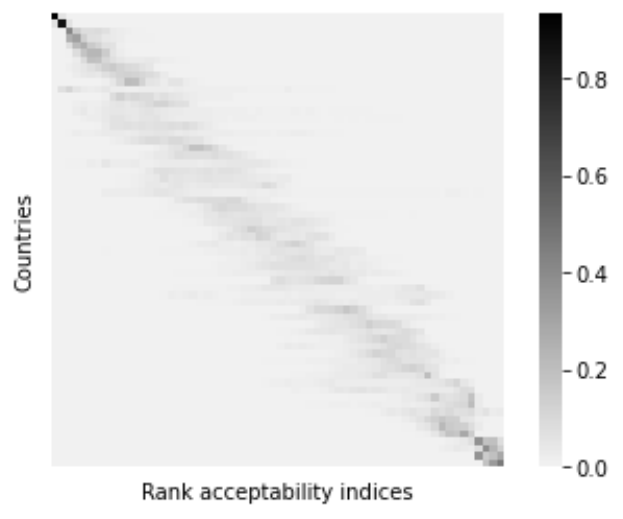}
\label{fig:smaa_ci_antigo_condorcet}}
\hfill
\caption{SMAA acceptability indices considering total absence of weight information.}
\label{fig:smaa_random}
\end{figure}

\begin{figure}[H]
\begin{centering}
\includegraphics[width=0.95\textwidth]{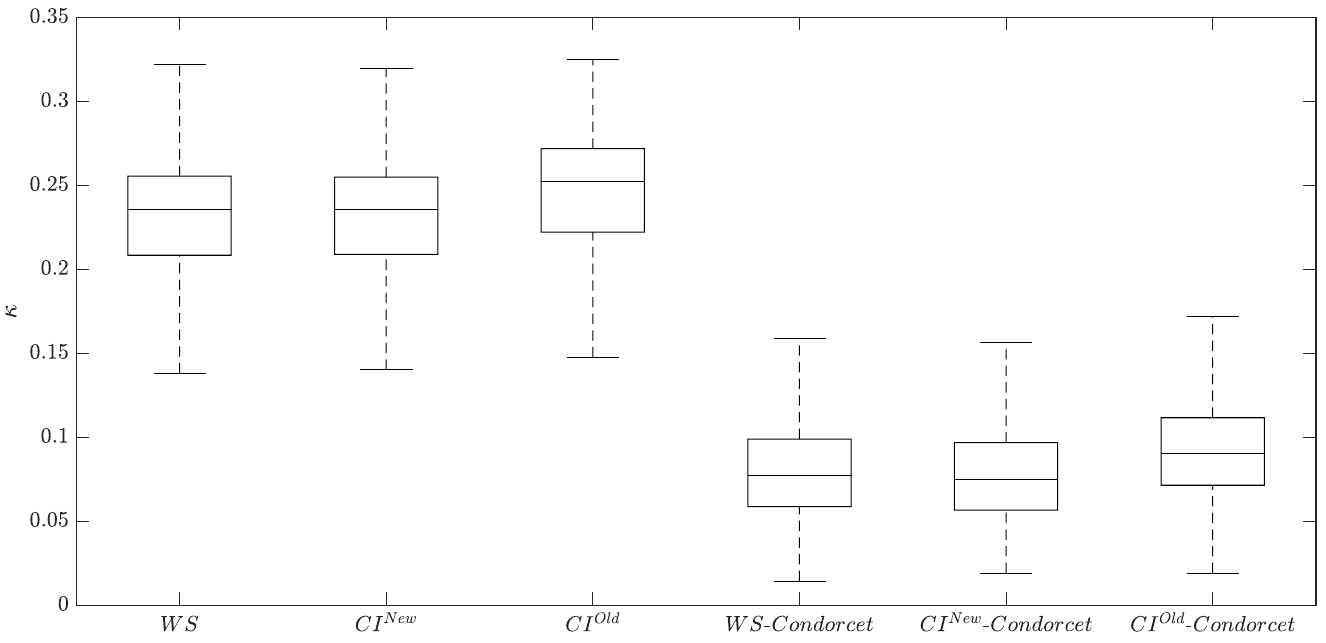} 
\par\end{centering}
\centering{}\caption{Visualization of variability in Kendall tau's distance of the different rankings considering total absence of weight information.
\label{fig:bloxplotrandom}}
\end{figure}

In order to analyze robustness in rankings, we elaborate two more results shown in Figures~\ref{fig:smaa_random} and~\ref{fig:bloxplotrandom}. In Figures \ref{fig:smaa_ws}, \ref{fig:smaa_ci}, and \ref{fig:smaa_ci_antigo}, we arrange the countries on the y-axis based on the GAII ranking and illustrate the acceptability indices of each country in the columns, obtained using SMAA with weighted sum, $CI^{u2}$, and $CI^{u1}$, respectively. In Figures \ref{fig:smaa_ws_condorcet}, \ref{fig:smaa_ci_condorcet}, and \ref{fig:smaa_ci_antigo_condorcet}, we organize the countries on the y-axis according to $WS$ - Cond., $CI^{u2}$ - Cond., and $CI^{u1}$ - Cond., respectively, and depict the acceptability indices of each country in the columns obtained using SMAA.

A primary observation regarding Figure \ref{fig:smaa_ws} concerns the high dispersion of the acceptability indices for countries in relation to the GAII ranking. This observation suggests that, in the total absence of weight information, the GAII does not exhibit robustness, as the probabilities of most countries being in their positions in this ranking are low. Conversely, the comparison between the acceptability indices and the $WS$ - Cond. ranking, illustrated in Figure \ref{fig:smaa_ws_condorcet}, shows less dispersion around their position in this ranking. A similar observation can be made by comparing Figure \ref{fig:smaa_ci} with Figure \ref{fig:smaa_ci_condorcet}, and Figure \ref{fig:smaa_ci_antigo} with Figure \ref{fig:smaa_ci_antigo_condorcet}.

The additional robustness analysis is illustrated in Figure \ref{fig:bloxplotrandom}. Each box in this figure represents the distribution of Kendall tau distances when comparing a specific ranking with the various rankings obtained from the SMAA simulation by randomly varying criteria weights. Therefore, the first box measures the Kendall tau dispersion when comparing the GAII ranking with all rankings obtained from SMAA. The second and third boxes indicate the Kendall tau dispersion by comparing the rankings presented in Table~\ref{tab:ranking}, which were obtained using the same weight values $w_j$ used to derive GAII, but applying $CI^{u2}$ and $CI^{u1}$, respectively, with all rankings obtained from SMAA. Finally, the last three boxes were obtained by comparing $WS$ - Cond., $CI^{u2}$ - Cond., and $CI^{u1}$ - Cond., respectively, with all rankings obtained from SMAA.

Upon analyzing Figure \ref{fig:bloxplotrandom}, one can observe that the three rankings—GAII, $CI^{u1}$, and $CI^{u2}$—performed less effectively than $WS$ - Cond., $CI^{u2}$ - Cond., and $CI^{u1}$ - Cond., as indicated by the the median Kendall tau, which were around 0.25 and 0.1, respectively. Furthermore, GAII, $CI^{u1}$, and $CI^{u2}$ exhibited a higher degree of dispersion when compared to $WS$ - Cond., $CI^{u2}$ - Cond., and $CI^{u1}$ - Cond. From this analysis, it is possible to infer that, in situations where it is not possible or fair to determine weight values, our approach obtains a more robust ranking.

\subsection{Robustness analysis of weights applying SMAA and Condorcet considering preference order of weights as defined in Tortoise}
\label{ssec:resultados_pesos_ordenado}

 In their report, Tortoise underscores the strategy employed in assigning weights to criteria. They argue that varying levels of data completeness across different sources necessitate careful integration of completeness considerations into the weighting system. Simultaneously addressing missing values through imputation, the weighting of indicators is adjusted in cases of limited data availability. Tortoise justifies reducing the relative weight of the indicator in such instances, emphasizing the importance of confidence in the data's representativeness, thereby allowing for a more substantial weighting of this factor.

 While the emphasis on data completeness is commendable, our prior study~\cite{BRACIS2023} demonstrated that the deterministic assignment of weights, without a comprehensive sensitivity analysis, can result in an indicator lacking robustness. The failure to consider the potential consequences of minor variations in weights undermines the stability and reliability of the indicator.

 To illustrate the sensitivity of the indicator to variations in weights, consider a scenario in which all weights remain constant except for $w_1$ (set at 0.11 by Tortoise) and $w_2$ (set at 0.06 by Tortoise). A marginal adjustment to 0.12 for $w_1$ and 0.05 for $w_2$, resulted in a shift in country rankings. Specifically, Canada and South Korea exchanged positions, underscoring the vulnerability of the indicator to minor alterations in weights.

 Therefore, in this section, we adhere to a preference weight order established by Tortoise to underscore the significance of confidence in the data's representativeness. However, we propose a more robust approach to derive the final ranking, aiming to mitigate sensitivity to minor variations in the weights. Additionally, we investigate the impact of data correlation on the final ranking to address potential biases in data aggregation.

In the initial analysis, we generated Table~\ref{tab:ranking_condorcet_ordered}, which illustrates the GAII ranking alongside our approaches ($WS$ - Cond., $CI^{u2}$ - Cond., and $CI^{u1}$ - Cond.), adhering to the preference weight order as in GAII. Examining the rankings presented in Table~\ref{tab:ranking_condorcet_ordered}, it is evident that the USA, China, Singapore, and the UK consistently hold the 1st, 2nd, 3rd, and 4th positions, respectively across all approaches. This consistency indicates their robustness to weight variations and aggregation methods, showcasing stability in the rankings.

 \begin{table}[ht]
\centering
\caption{Comparison of ranking positions for the top ten countries in GAII and those determined using our approaches, considering preference order of weights.}
\begin{tabular}{lcccc}
\multicolumn{1}{c}{\multirow{2}{*}{Position}} & \multicolumn{4}{c}{Countries}                                                                          \\ \cline{2-5} 
\multicolumn{1}{c}{}                          & GAII                                          & $WS$ -  Cond. & $CI^{u2}$ - Cond. & $CI^{u1}$ -  Cond. \\ \hline
1st                                           & USA                                           & USA           & USA               & USA                \\
2nd                                           & China                                         & China         & China             & China              \\
3rd                                           & Singapore                                     & Singapore     & Singapore         & Singapore          \\
4th                                           &  UK      & UK            & UK                & UK                 \\
5th                                           & \cellcolor{Peach} Canada     &  Israel        & \cellcolor{Butter} Switzerland       & \cellcolor{Butter} Switzerland        \\
6th                                           & South Korea & \cellcolor{Butter} Switzerland   & \cellcolor{Peach} Canada            & Germany            \\
7th                                           &  Israel                                        & \cellcolor{Peach} Canada        & Israel            & South Korea        \\
8th                                           &  Germany   & Germany       & Germany           &  Israel             \\
9th                                           & \cellcolor{Butter} Switzerland                                   &  South Korea   &  South Korea       & \cellcolor{Peach} Canada             \\
10th                                          & Finland                                       & Netherlands   & Netherlands       & India              \\ \hline
\end{tabular}
\label{tab:ranking_condorcet_ordered}
\end{table}

 Conversely, some countries undergo changes in their positions when considering different approaches. For instance, Switzerland, which holds the 7th position in GAII, rises to the 6th position in $WS$ - Cond., and 5th in  $CI^{u2}$ - Cond. and $CI^{u1}$ - Cond. Furthermore, Canada is ranked 5th in GAII, while in $WS$ - Cond., it holds the 7th position; in $CI^{u2}$ - Cond., it is placed 6th, and in $CI^{u1}$ - Cond., it falls to the 9th position. This variability highlights the sensitivity in their positions, indicating potential changes when adjusting weights or considering correlations. 

To identify disparities between these rankings, we calculate the Kendall tau distances obtained through pairwise comparisons of the complete rankings across all countries. We find that this distance between GAII and $WS$ - Cond. is 0.26, between GAII and $CI^{u2}$ - Cond. is 0.2, and between GAII and $CI^{u1}$ - Cond. is 0.31. Additionally, the $\tau$ value between $WS$ - Cond. and $CI^{u2}$ - Cond. is 0.18. This result underscores that the rankings obtained from different approaches differ when weights are not deterministic but follow a preference order, as well as when considering biases in the data.

\begin{figure}[h]
\centering
\subfloat[$WS$ - GAII's ranking.]{\includegraphics[width=2.5in]{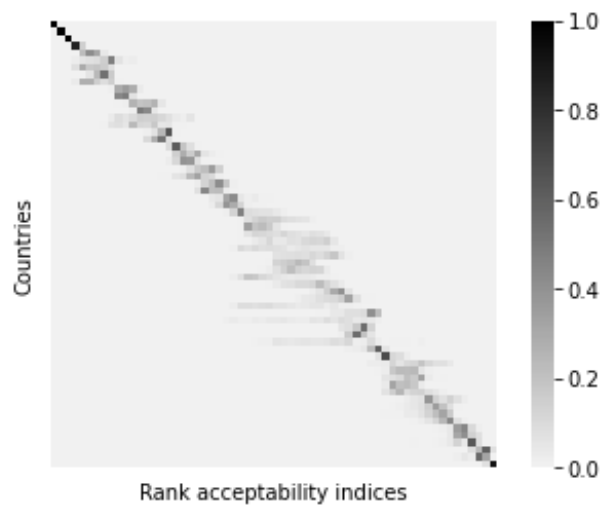}
\label{fig:smaa_ordered_ws}}
\hfill
\subfloat[$WS$ - Condorcet's ranking.]{\includegraphics[width=2.5in]{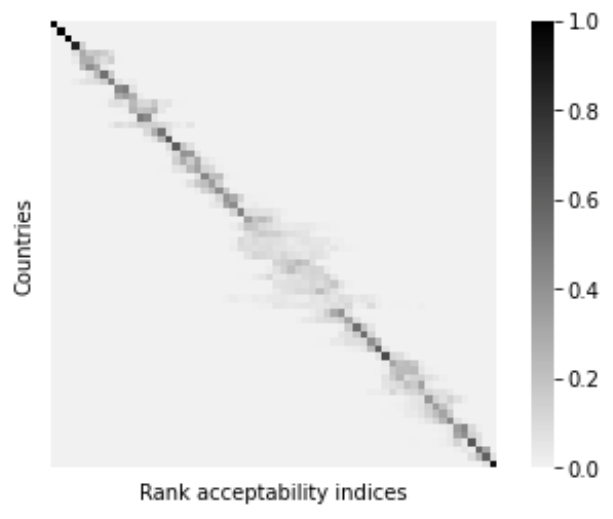}
\label{fig:smaa_ordered_ws_condorcet}}
\hfill
\subfloat[$CI^{u2}$ - Condorcet's ranking.]{\includegraphics[width=2.5in]{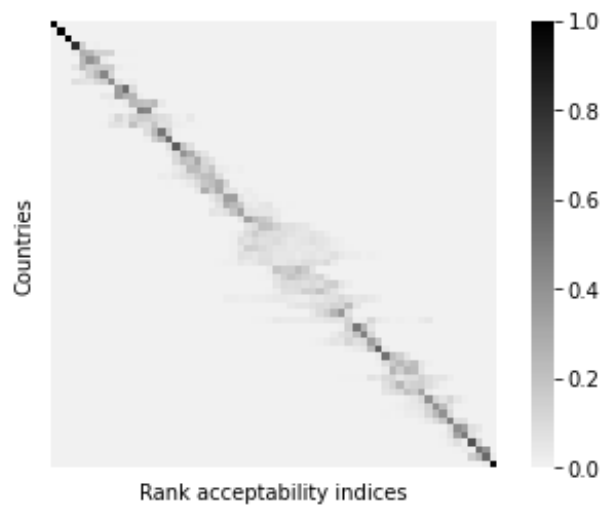}
\label{fig:smaa_ordered_ci_condorcet}}
\hfill
\subfloat[$CI^{u1}$ - Condorcet's ranking.]{\includegraphics[width=2.5in]{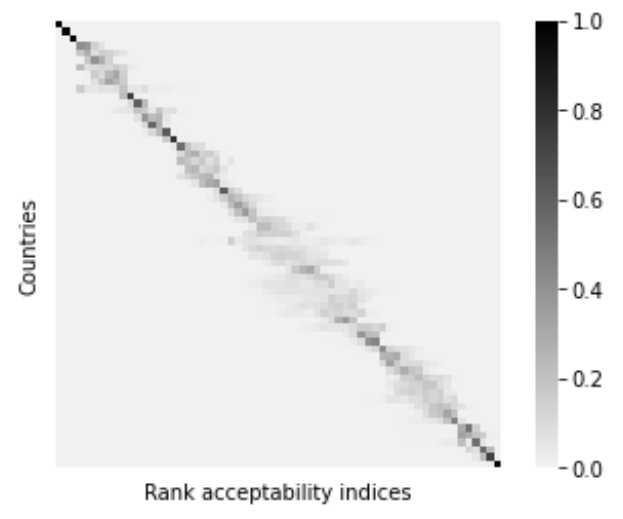}
\label{fig:smaa_ordered_ci_antigo_condorcet}}
\hfill
\caption{SMAA Acceptability indices considering preference order of weights.}
\label{fig:smaa_ordered_random}
\end{figure}

\begin{figure}[h]
\begin{centering}
\includegraphics[width=0.95\textwidth]{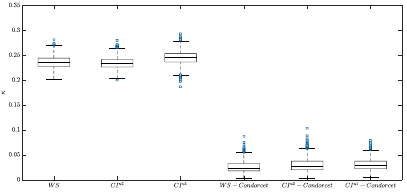} 
\par\end{centering}
\centering{}\caption{Kendall tau's distance for ordered weights.
\label{fig:bloxploordered}}
\end{figure}

 The next step in the analysis is to determine which ranking is more robust in terms of changes in the weights' values. For this analysis, we elaborate Figures~\ref{fig:smaa_ordered_random} and~\ref{fig:bloxploordered}. In Figure \ref{fig:smaa_ordered_ws}, we arrange the countries on the x-axis based on the GAII ranking and illustrate the acceptability indices of each country in the columns, obtained using SMAA plus weighted sum, while adhering to the same preference order as in GAII. This figure illustrates the probability of the countries being in the positions as in GAII. In Figures \ref{fig:smaa_ordered_ws_condorcet}, \ref{fig:smaa_ordered_ci_condorcet}, and \ref{fig:smaa_ordered_ci_antigo_condorcet}, we organize the countries on the y-axis according to $WS$ - Cond., $CI^{u2}$ - Cond., and $CI^{u1}$ - Cond., respectively, and depict the acceptability indices of each country in the columns obtained using SMAA. These figures indicate the probability of the countries being in the positions as in $WS$ - Cond., $CI^{u2}$ - Cond., and $CI^{u1}$ - Cond. rankings.

The comparison between Figures \ref{fig:smaa_ordered_ws} and \ref{fig:smaa_ordered_ws_condorcet} indicates that the ranking $WS$ - Cond. is more robust, since the probability of a certain country be in their position in the ranking is high in most of countries, regardless the weight associated to the criteria. Instead, in GAII, one can note that these probabilities are more scattered. In addition, some countries present a great probability of being in a position that is not their position in GAII. Similar results can be found in the comparison between Figures \ref{fig:smaa_ordered_ws} and \ref{fig:smaa_ordered_ci_condorcet}, and Figures \ref{fig:smaa_ordered_ws} and \ref{fig:smaa_ordered_ci_antigo_condorcet}.

To delve deeper into the robustness analysis, Figure~\ref{fig:bloxploordered} illustrates the distribution of the Kendall tau distances when comparing a specific ranking with various rankings obtained from SMAA simulations by adjusting criteria weights in the order presented in GAII. The first box in this figure quantifies the dispersion when comparing the GAII ranking with all rankings obtained from SMAA, following the same weight order as in GAII. Similarly, the second and third boxes indicate the distance dispersion by comparing the rankings presented in Table~\ref{tab:ranking}. These rankings were obtained using the same weight values $w_j$ used to derive GAII but applying $CI^{u2}$ and $CI^{u1}$, respectively, with all rankings obtained from SMAA. Finally, the last three boxes compare $WS$ - Cond., $CI^{u2}$ - Cond., and $CI^{u1}$ - Cond., respectively, with the rankings obtained from SMAA using their respective methods.

By comparing Figure~\ref{fig:bloxploordered} with Figure~\ref{fig:bloxplotrandom}, which was generated using completely random weights, it becomes apparent that all boxes in Figure~\ref{fig:bloxploordered} exhibit lower dispersion. This outcome suggests that, when adhering to a preference order, the rankings demonstrate better convergence, i.e., less variation when weights are altered.

Upon analyzing Figure \ref{fig:bloxploordered}, a similar trend emerges as when weights are generated entirely randomly in SMAA. In other words, the three rankings—GAII, $CI^{u1}$, and $CI^{u2}$—exhibited less effectiveness compared to $WS$ - Cond., $CI^{u2}$ - Cond., and $CI^{u1}$ - Cond. Our proposed rankings, derived from $WS$ - Cond., $CI^{u2}$ - Cond., and $CI^{u1}$ - Cond., demonstrate significant robustness, as indicated by the average Kendall tau close to zero. This result implies that these rankings are less sensitive to changes in weights. Thus, with this finding, we demonstrate that even when a preference order is required to ensure completeness across various data sources, a more robust ranking can be achieved without assigning deterministic values to the weights.

\section{Conclusion}
\label{sec:conclusions}

 This paper presented an analysis of AI indicators for comparing countries. Our objective was to address three questions raised in the introduction: the first concerning the hypothesis of interactions between AI dimensions, the second regarding the influence on the ranking when changing criteria weights, and the last one concerning the determination of a robust AI ranking that does not require consideration of deterministic weights. To answer these questions, we applied the weighted sum with SMAA and the Choquet integral with SMAA to analyze the Tortoise GAII in terms of criteria weights and aggregation procedures. Additionally, we propose three approaches, namely $WS$ - Cond., $CI^{u2}$ - Cond., and $CI^{u1}$ - Cond., to derive robust rankings that do not rely on deterministic weights. Moreover, the $CI^{u2}$ - Cond. and $CI^{u1}$ - Cond. approaches also consider the interaction between criteria.

 Regarding the first issue, the results confirm significant correlations among AI dimensions. We observe that criterion 5 (Research) demonstrates high correlations with other criteria. However, in establishing the GAII ranking, it was assigned the criterion associated with the highest weight value. Consequently, the criterion that should have had a higher penalization to prevent bias was the one with greater relevance. Furthermore, our proposal to use the Choquet integral revealed that rankings can be very different, highlighting the relevance of applying this method to mitigate bias.

 With respect to the second and third questions, we employed two distinct strategies. In the first one, we generated weights in SMAA entirely randomly, while in the second, these weights were randomly generated but followed an ordinal arrangement identical to that used to obtain GAII. In a sensitivity analysis for the first strategy, we observed a significant difference between the GAII ranking, which utilized deterministic weights, and our rankings without deterministic weights. For example, the UK, occupying the 4th position in GAII, does not exhibit the highest probability of being in the 4th position when weights are not deterministic, and it drops to the 7th position in our rankings. 
 
  The sensitivity analysis also indicated that the ranking $WS$ - Cond. is more robust in terms of acceptability indices, demonstrating a higher probability for countries to maintain their positions in the $WS$ - Cond. ranking compared to GAII. Additionally, a the comparison in terms of Kendall tau distance of each ranking with the corresponding rankings obtained by altering weights showed that, on average, our rankings are more closely aligned with those resulting from weight variations than GAII. This suggests a reduced sensitivity to changes in weights when use our proposal.

 Concerning to the second strategy, where weights were randomly generated but followed an ordinal arrangement, our finds were similar to the first one. Results showed the GAII to be sensitive to weights, and some countries are in certain position on GAII but with higher probability to be in a different position. Our proposal rankings showed to be more robust in terms of the rank acceptability indices. Finally, in terms of Kendall tau distance, our proposal rankings showed to be, in average, very close to the rankings obtained in SMAA by varying weights.

 Thus, some advantages in the ranking $WS$ - Cond. are related to reducing subjectivity in weight choice and enhancing robustness, leading to a higher probability of a country maintaining its position. Concerning the first strategy, which involves using completely random weights, the advantage lies in the ease of comparing the position of each country over the years, as the values of the weights do not change. However, this strategy has the disadvantage of not considering the relative importance of criteria, and it overlooks data reliability and  expert opinions. This limitation can be alleviated by implementing the second strategy, which enables establishing an order of preference among criteria without the necessity of determining specific values, thereby contributing to the reduction of subjectivity.

 \bibliographystyle{elsarticle-num} 
 \bibliography{cas-refs}





\end{document}